\documentstyle[11pt,aaspp4]{article}

\received{1998 March 4}

\begin{document}     

\def\today{\ifcase\month\or
January\or February\or March\or April\or May\or June\or
July\or August\or September\or October\or November\or December\fi
\space\number\day, \number\year}

\baselineskip 0.175in

\newcommand{\squig}{$\sim$}
\newcommand{\squigleq}{\mbox{$^{<}\mskip-10.5mu_\sim$}}
\newcommand{\squiggeq}{\mbox{$^{>}\mskip-10.5mu_\sim$}}
\newcommand{\squiggeqmm}{\mbox{$^{>}\mskip-10.5mu_\sim$}}
\newcommand{\decsec}[2]{$#1\mbox{$''\mskip-7.6mu.\,$}#2$}
\newcommand{\decsecmm}[2]{#1\mbox{$''\mskip-7.6mu.\,$}#2}
\newcommand{\decdeg}[2]{$#1\mbox{$^\circ\mskip-6.6mu.\,$}#2$}
\newcommand{\decdegmm}[2]{#1\mbox{$^\circ\mskip-6.6mu.\,$}#2}
\newcommand{\decsectim}[2]{$#1\mbox{$^{\rm s}\mskip-6.3mu.\,$}#2$}
\newcommand{\decmin}[2]{$#1\mbox{$'\mskip-5.6mu.$}#2$}
\newcommand{\asecbyasec}[2]{#1$''\times$#2$''$}
\newcommand{\aminbyamin}[2]{#1$'\times$#2$'$}

\title{A Technique for Narrowband Time Series Photometry:
the X-ray Star V2116~Oph}
\author{Eric W. Deutsch and Bruce Margon}
\affil{Department of Astronomy, 
       University of Washington, Box 351580,
       Seattle, WA 98195-1580\\
       deutsch@astro.washington.edu; margon@astro.washington.edu}

\author{and\\ \vskip .1in Joss Bland-Hawthorn}
\affil{Anglo-Australian Observatory,
       P.O. Box 296, Epping, NSW 2121, Australia\\
       jbh@aaossz.aao.gov.au}

\begin{center}
Accepted for publication in PASP\\
To appear in volume 110, August 1998\\
{\it received 1998 March 4}; \ {\it accepted 1998 May 25}
\end{center}

\begin{abstract}

We have used innovative features of the Taurus Tunable Filter
instrument on the 3.9-m Anglo-Australian Telescope to obtain
nearly-continuous, high-throughput, linear photometry of V2116~Oph in a
7 \AA\ bandpass at the center of the O\,I $\lambda$8446 emission line.
This instrumental technique shows promise for applications requiring
precise, rapid, narrowband photometry of faint objects.

The spectrum of V2116~Oph, the counterpart of GX\,1+4 (=X1728--247), is
exotic, even among the unusual spectra of other optical counterparts of
compact Galactic X-ray sources.  The second strongest emission line is
an unusual one, namely extremely prominent O\,I $\lambda$8446, which is
likely to result from pumping by an intense Ly$\beta$ radiation field.
As the X-radiation from GX\,1+4 is steadily pulsed, with typical pulsed
fractions of 0.4, the O\,I~$\lambda$8446 emission in V2116~Oph may also
be strongly modulated with the current 127~s period of the X-ray
source.  If so, this may well allow us to obtain high signal-to-noise
radial velocity measurements and thus to determine the system
parameters.  However, no such pulsations are detected, and we set an
upper limit of $\sim1$\% (full-amplitude) on periodic $\lambda$8446
oscillations at the X-ray frequency.  This value is comparable to the
amplitude of {\it continuum} oscillations observed on some nights by
other workers.  Thus we rule out an enhancement of the pulsation
amplitude in O\,I emission, at least at the time of our observations.

\end{abstract}


\clearpage
\section{INTRODUCTION}

GX\,1+4 (=X1728--247), a classical, luminous X-ray binary observed for
25 years, is projected very close to the galactic center and probably
is at a distance of $\sim8$~kpc, as the inferred X-ray luminosity is
then $6\times10^{37}$ erg~s$^{-1}$, near the Eddington limit.  The
spectrum of the $V\sim19$ optical counterpart, first identified by
Glass \& Feast (1973) and now known as V2116~Oph, has at times shown
higher excitation emission lines than any other known X-ray star; for
example, [Fe\,X] (I.~P. = 235 eV) and [Ar\,XI] appeared (Davidsen et al.
1977).  The spectrum appears to be markedly time variable on scales
from minutes to years; in recent epochs, the highest excitation lines
have disappeared (Chakrabarty \& Roche 1997), but enormously strong
H$\alpha$ emission remains.

The symbiotic-like optical spectrum of V2116~Oph directly shows the
presence of a red giant, of type near M5\,III (Chakrabarty \& Roche
1997), and suggests that we may be viewing the system at a very
special, short-lived stage, when the normal primary is passing through
a quite brief phase of its evolution.  This point is made even more
vivid by the X-ray behavior.  GX\,1+4 is an X-ray pulsar with a
coherent X-ray period of about 130~s, a pulse amplitude \squig 0.4, and
an enormous X-ray period {\it derivative} of $\dot P =
dP/dt\sim-3$~s~yr$^{-1}$ (Laurent et al. 1993 and references therein).
While the period is slow, although not inordinately slow for X-ray
binaries, the spin-up rate is the fastest for any known X-ray pulsar.
The characteristic age, $P/\dot P \sim40$~yr, confirms that this is an
amazingly rapidly evolving object.  More recent X-ray observations show
that the X-ray $\dot P$ has reversed sign, although the modulus remains
very large (Laurent et al. 1993, Chakrabarty et al. 1997).  It seems
clear that in GX\,1+4 we have the chance to observe an X-ray pulsar
undergoing rapid evolution.

Yet even the orbital parameters of this system remain unknown, and the
detection of an optical analogue to the X-ray pulses could yield an
elegant and accurate radial velocity solution (\S 3).  Recently
broadband optical flickering and pulsations at the X-ray period have
indeed been reported from V2116~Oph, at an amplitude of a few percent
(Jablonski et al. 1997, Jablonski \& Pereira 1997).  These data seem to
indicate a complex and/or erratic dependence of pulse amplitude on
wavelength, optical brightness of the system, etc., and further
observations will clearly be needed to sort out the situation.

The second strongest emission line after H$\alpha$ in the optical
spectrum of V2116~Oph is a remarkable one, namely extremely prominent
O\,I $\lambda$8446, which can be readily seen in our low-resolution
spectrum in Fig. 1.  This line has been reported in a small number of
interesting objects ranging from Seyfert galaxies to occasional
odd stars.  In some objects, the great observed strength relative to
other common species must be explained by some type of preferential
emission mechanism.  Grandi (1975, 1976) suggested that pumping by an
intense Ly$\beta$ radiation field, in a wavelength coincidence
(Ly$\beta\ \lambda1025.72\approx {\rm O\,I} \ \lambda1025.76$) not
dissimilar to the famous Bowen mechanism in which He II Ly $\alpha$
pumps O\,III and N\,III in nebulae and mass-exchange binaries, may be
responsible.  O\,I $\lambda$8446 has been reported at great strength in
the symbiotic star V1016~Cyg (Rudy et al. 1990), and the presence there
of the OI $\lambda11287$ line at the expected strength confirms that
Ly$\beta$ pumping is the mechanism.

As V2116 Oph is exposed to $\sim10^{38}$ erg s$^{-1}$ of pulsed ionizing
radiation, it seems possible that pulsed Lyman photons may trigger coherent
127~s pulsations in O\,I $\lambda$8446, possibly at quite large amplitudes.
Here we report on a search for these pulsations using an innovative
technique with applicability to a variety of other problems.

\section{OBSERVATIONS AND DATA REDUCTION}

The Taurus Tunable Filter (TTF) is a narrowband interference filter
consisting of a red ``arm'' covering 6300~\AA~$-$~9600~\AA\ and a blue
``arm'' covering 3700~\AA~$-$~6500~\AA, with an adjustable passband of
between 6 \AA\ and 60 \AA.  It is now available for use on the
Anglo-Australian Telescope (AAT) and William Herschel Telescope and is
described in detail by Bland-Hawthorn \& Jones (1998).  Frequency
switching of the etalon can be synchronized with the movement of charge
(charge shuffling) up or down the CCD columns.  When combined with an
arbitrary mask, the instrument can expose only part of the CCD, leaving
the rest to act as a charge storage device.  This provides many
different observing configurations of narrowband imaging, applicable to
studying a wide range of astrophysical problems.

An elegant way to obtain continuous, high-throughput, linear photometry
in a narrow bandpass at arbitrary wavelength is to use the
charge-shuffle mode of the TTF.  Here we discuss an attempt
at the AAT to determine if O\,I $\lambda$8446 pulsations in V2116~Oph
can be used to obtain high signal-to-noise radial velocity
measurements and thus to elucidate the system parameters.

On 1997 July 10, we repeatedly imaged V2116~Oph using the TTF
on the AAT through a $2' \times 6''$ slit, rotated to position angle
64$^\circ$ to include the X-ray star plus one brighter and one fainter
nearby companion.  Figure 2 shows a \aminbyamin{3}{3} region near
V2116~Oph and includes the outline of this slit superposed.  To tune
the TTF bandpass to the center of the emission line, we step the etalon
in wavelength across the line.  Figure 3 shows the result of the scan
and demonstrates that the O\,I $\lambda$8446 emission line is
well-detected; we set the etalon to step 180 and a width of 7\,\AA.  In
this configuration, TTF yields 90\% filter transmission, and we
achieved \squig 500 detected photons s$^{-1}$ during cloudbreaks.

Each slit image is then exposed for 12~s before charge is shuffled 10
pixels (6$''$) down the CCD.  This is repeated 102 times before the
chip becomes full and is read out, a quite efficient protocol as each
shuffle consumes only 1~s, a delay time selected to avoid burning out
the shutter.  There is only about a 100~s delay between the start of a
new series of exposures and the end of the previous series, as the CCD
is read out and written to disk.  The virtues of this technique over
alternatives such as time-resolved dispersive spectroscopy include
significantly higher throughput (important given the narrow wavelength
band of interest in this faint object) and excellent photometric
accuracy due to the lack of a narrow slit.

A new mode of the instrument, not available at the time of these
observations, now allows the charge to be read out as it is shuffled
down the chip, eliminating the need to read the entire chip at once and
the time lost while doing so.  In addition, the charge can be shifted
quickly, eliminating the need to close and open the shutter between
each exposure.  The shuffle speed is about 50 $\mu$s per row, and the
read out speed is about 40 ms per row.  These new modes now allow TTF
to obtain a truly continuous time series dataset.

Figure 4 displays the five separate exposures (each totaling 20~min of
integration) obtained on the partly-cloudy night of
\decsec{2}{5}$-$\decsec{3}{0} seeing.  Within each strip, V2116 Oph is
the bright object to the right; the brighter reference star ({\it a}) is
the object on the left of each strip.

Each image is first inspected and manually cleaned of charged-particle
hits, and then aperture photometry is used to measure a magnitude for
the program and reference stars in each subframe.  In this case
aperture photometry performed better than profile fitting software, as
poor seeing conditions caused the stellar profile to vary considerably
between the short exposures.  Due to the cloudy conditions, the direct
measured magnitudes vary considerably, but differential photometry
effectively removes this effect and allows us to achieve \squig2\%
relative photometry on V2116 Oph during times of average obscuration.
From these measurements we derive a light curve which is nearly
continuous over 2.1 hr.  The standard deviation is consistent with that
expected from count-rate uncertainties and shows no obvious periodic
or aperiodic variation.  The power spectrum of the best single series
of exposures reveals no significant power down to a $3\sigma$ level of 3\%
(full-amplitude) at the $126.971 \pm 0.005$ s period of GX\,1+4, as
determined by Chakrabarty (1997) using BATSE data from the week of our
TTF observations.  This upper limit is determined by adding artificial
sinusoidal oscillations (with proper counting and cloud obscuration
uncertainties added) and measuring when the signal becomes significant
in the power spectrum.  When the light curves from all five frames are
combined into a power spectrum and also averaged into 10 phase bins
using the BATSE period, no significant oscillations are detected to a
$3\sigma$ level of 1\% (full-amplitude).

\section{DISCUSSION}

With a giant primary and any reasonable assumption for the mass of the
secondary (presumably a neutron star), the system orbital period must
be of order one year.  Although we would very much like to
understand even the basic parameters of this unusual system, simply
extracting even the period will not be easy.  Searches for periodic
variations in the X-ray pulse timing residuals are foiled by large,
irregular torque derivatives (Chakrabarty et al. 1997, Bildsten et al.
1997).  Normal radial velocity spectroscopy of the optical lines will
be difficult; the expected variations are very small compared at least
to the very broad H$\alpha$ emission.  Searches for periodic variations
in the H$\alpha$ intensity (Greenhill et al. 1995) and H$\alpha$
profile (e.g., Sood et al. 1995) have not revealed a significant
orbital period component in the large variations, presumably due to
accretion rate fluctuations.

In principle the detection of an optical analogue to the X-ray pulses
could yield an accurate radial velocity solution.  If the
ionizing radiation is reprocessed to visible light at a location in the
system fixed with respect to the barycenter, and the recombination
times  are short compared to the pulse period, the resulting optical
pulses can be searched for periodic modulation due to orbital motion.
If the reprocessing surface is physically larger than the light travel
distance in one pulse cycle, phase mixing may dilute the pulse amplitude
greatly, but working in the Fourier domain provides far more potential
sensitivity than standard radial velocity spectroscopy.  This technique
has of course been applied decades ago, with great success, to the
X-ray pulsar HZ~Her/Her X-1 (Middleditch \& Nelson 1976).

We have conducted a first search for pulsations of O\,I $\lambda$8446
in GX\,1+4, in the hope of establishing if measurement of the periodic
modulation of this strong emission line may be used to obtain high
signal-to-noise radial velocity measurements, and thus to determine the
system parameters.  Despite less than optimal observing conditions, we
are able to rule out an enhancement of the pulsation amplitude in O\,I
at the time of our observations; our combined data do not reveal any
pulsation at the X-ray period, with a $3\sigma$ upper limit of 1\%
(full-amplitude).  Krzeminski \& Priedhorsky (1978) report similar
limits to H$\alpha$ periodic variations.  Chakrabarty et al. (1998)
report $2\sigma$ upper limits of 5\% and 9\% for pulsation in infrared
He I and Pa\,$\beta$, respectively, at the time of their observations.
Broadband optical pulsations up to $\sim5$\% have been observed, but
only on certain occasions, and broadband pulsation upper limits of
0.1\% have also been documented (Jablonski et al. 1997).  Those authors
have reported a correlation where these pulsations are stronger during
brighter optical states.  However, our data do not allow us to
determine the absolute brightness of V2116~Oph during the
observations.

The lack of pulsation enhancement in O\,I $\lambda$8446 allows several
possible interpretations.  If the Ly~$\beta$ photons responsible for
pumping the O\,I are indeed created by reprocessed X-rays, the
reprocessing region must be sufficiently large to completely dilute the
pulse amplitude.  A circumstellar nebula such as discussed by
Chakrabarty \& Roche (1997) is one such possibility.  A second possible
interpretation follows a suggestion of Chakrabarty \& Roche (1997)
based on other optical emission lines: the Ly~$\beta$ photons are
excited by the thermal emission of the accretion disk instead of
reprocessed X rays.

Substantial further observations of this complex system are needed to
unravel its nature.  Optical pulsations in HZ Her are well-known to be
detected only during a highly restricted subset of orbital and
precessional phases in the system.  Nonetheless, the unique utility of
TTF in charge shuffle mode for narrow band time-series photometry of
faint objects is clear.

\acknowledgments

We thank Deepto Chakrabarty for providing us with the pulse period for
the time of our observations from BATSE data, as well as the ATAC
for granting time for this project.


\clearpage

\begin{figure}
\plotone{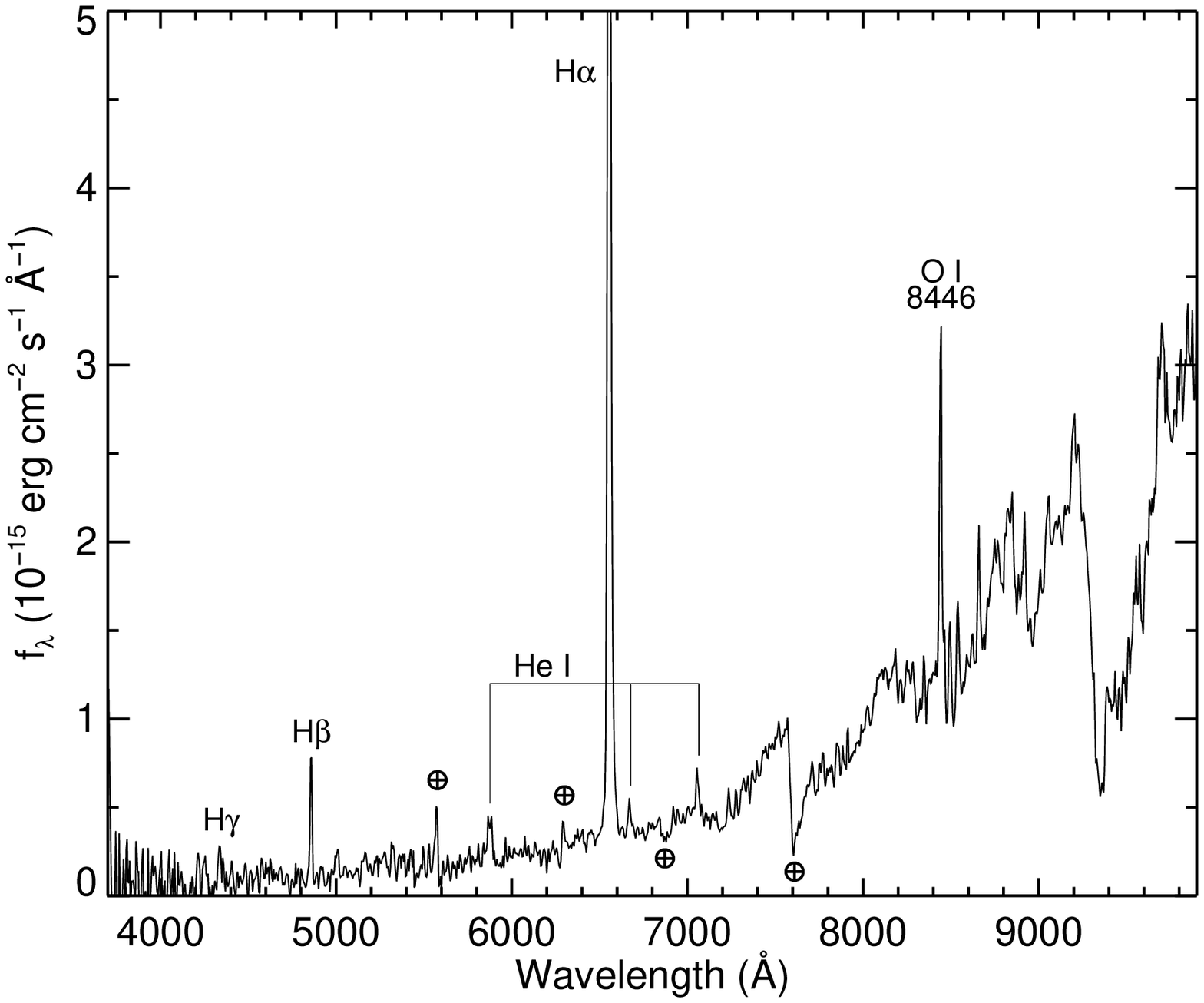}
\caption{Spectrum of V2116 Oph obtained on 1994 July 11 with the Apache
Point 3.5-m telescope and Dual Imaging Spectrograph in its \squig 20 \AA\
resolution mode.  Note the extraordinarily strong O\,I $\lambda8446$.
He I $\lambda\lambda5876,6678,7065$ are indicated; other features
are discussed in Chakrabarty \& Roche (1997).}
\end{figure}

\begin{figure}
\plotone{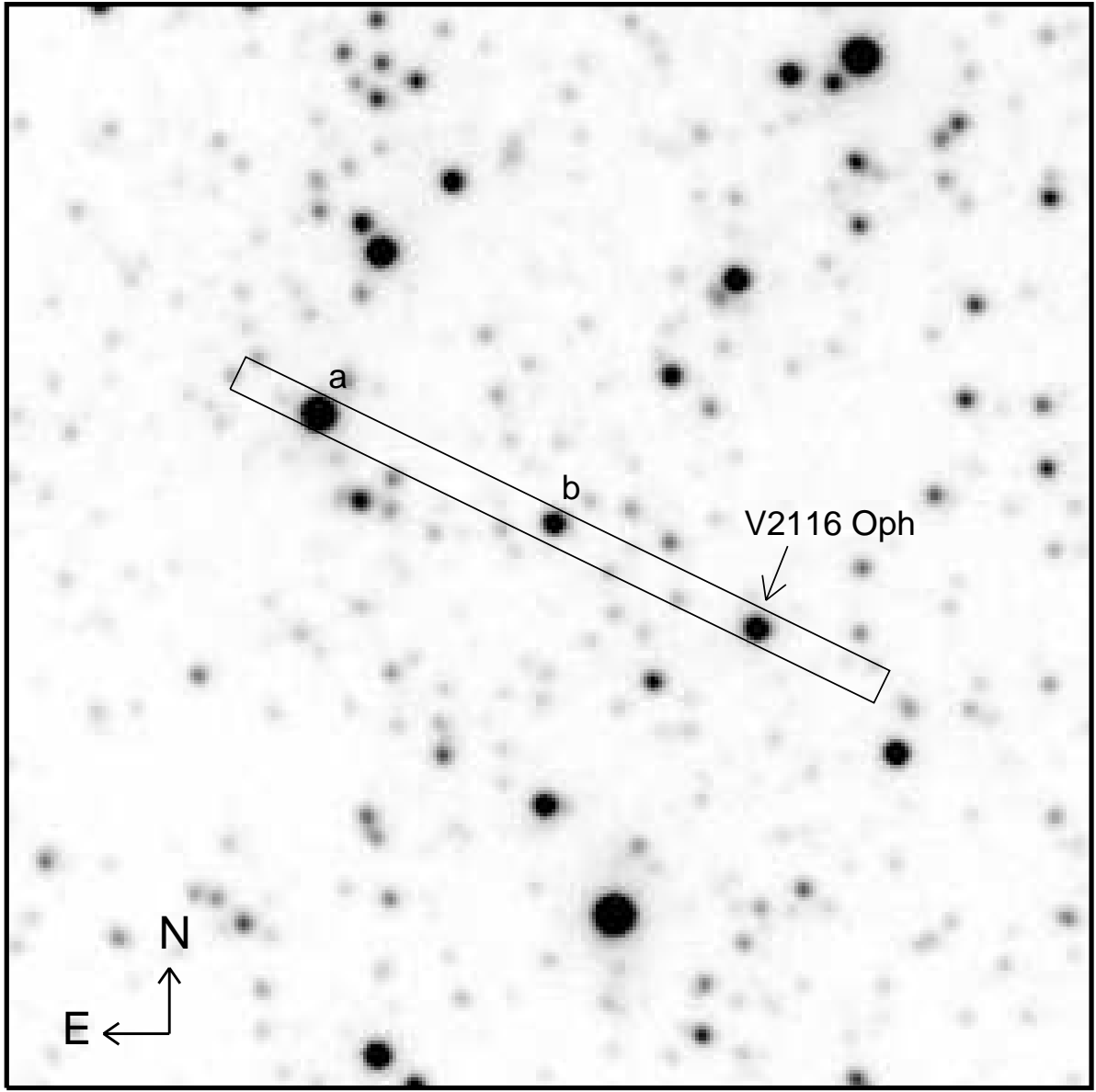}
\caption{\aminbyamin{3}{3} I-band image of V2116 Oph and the
surrounding region, obtained with the Apache Point 3.5-m telescope.
The $2' \times 6''$ exposure mask, rotated to position angle 64$^\circ$
as used in the AAT TTF time-series observations, is overlaid.  The two
bright stars (denoted $a$ and $b$) within the mask NE of the X-ray source
are the comparison stars that appear in the time-resolved photometric
data of Fig.~4.}
\end{figure}

\begin{figure}
\plotone{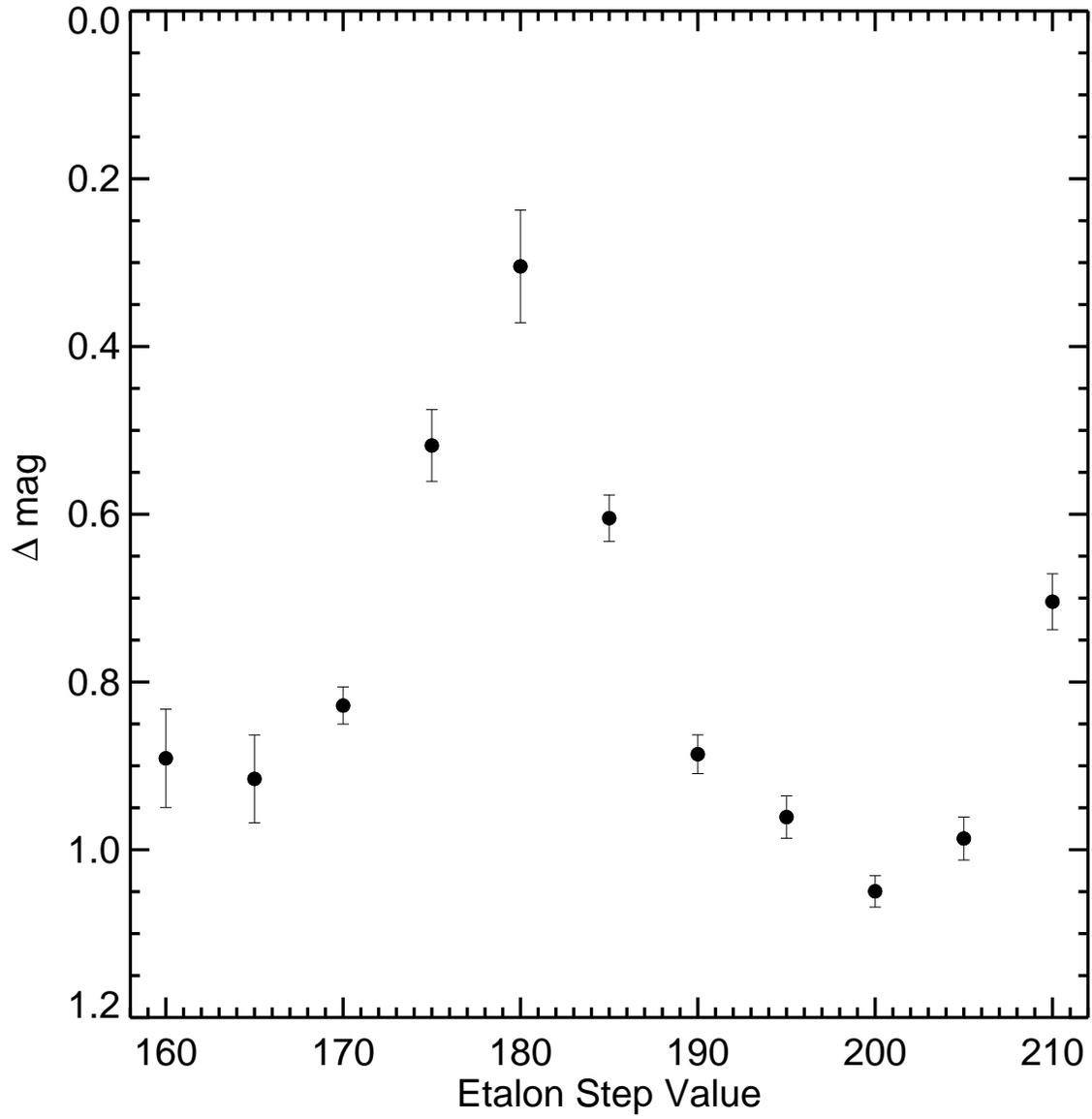}
\caption{Relative magnitude of V2116~Oph to a nearby reference star as
the TTF etalon is stepped in wavelength.  Each stepping interval is 5
etalon units or 2.8 \AA; the FWHM of the passband is 7 \AA.  The peak
corresponds to the O\,I $\lambda$8446 emission line, confirming that
the line is present during our time series observations.  The variable
size of the error bars reflects changes in cloud obscuration during the
scan.}
\end{figure}

\begin{figure}
\plotone{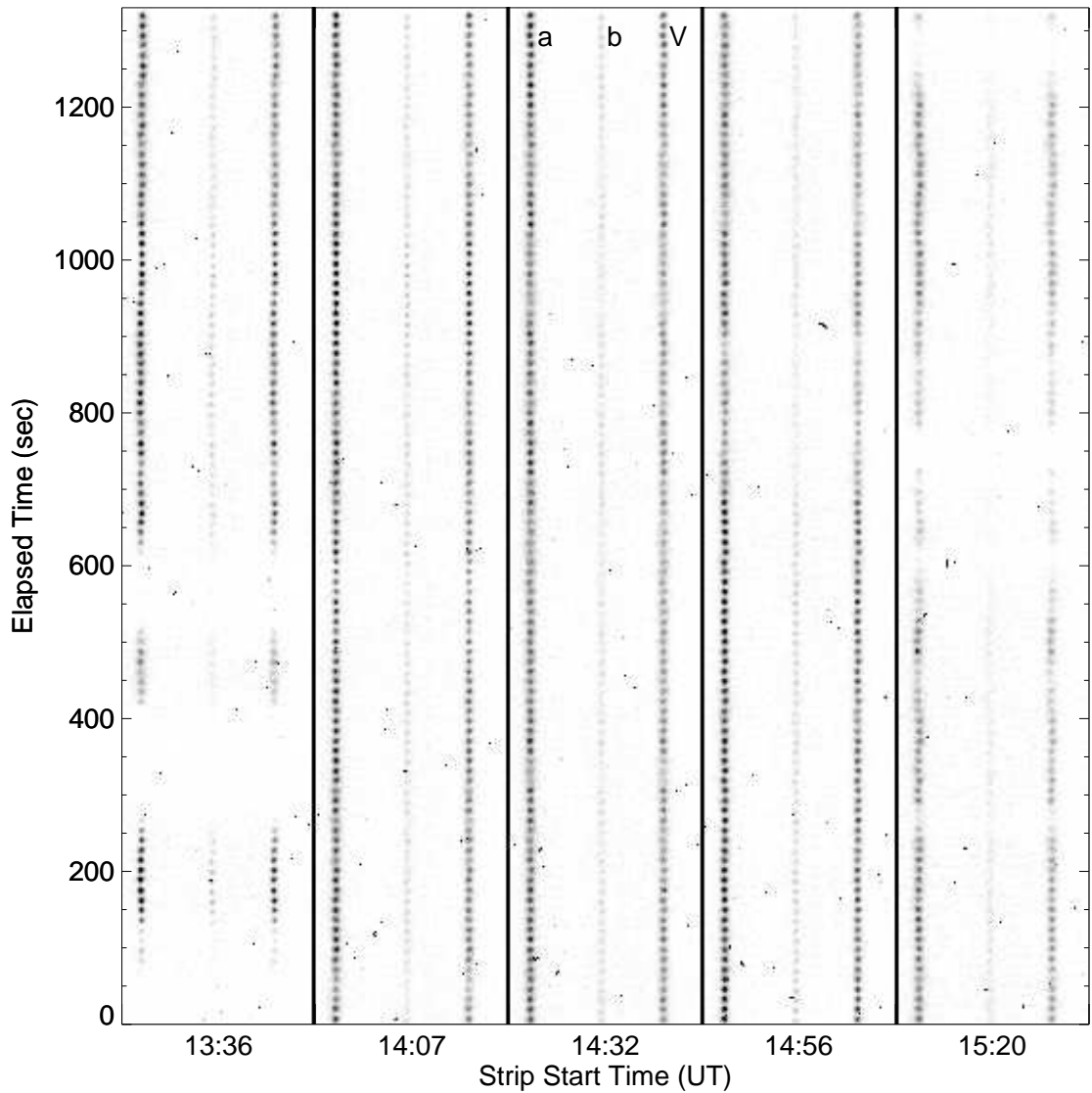}
\caption{Five strips of 102 $2' \times 6''$ 12-second exposures of the
X-ray star V2116~Oph ($V$) and two nearby reference stars ($a$ and $b$) made
with the TTF set to a 7 \AA\ bandpass centered on O\,I $\lambda$8446.
Time increases vertically in each strip (note the greater frequency of
charged-particle hits at the bottom of each image, where charge has
been on the CCD the longest).  The obvious variations in intensity are
due to variable cloud obscuration.  During the first and last strip,
cloud obscuration was sufficient that autoguiding had to be switched
off, as can be seen from the spatial oscillation in the images.}
\end{figure}

\end{document}